\documentclass[12pt,preprint]{aastex}
\usepackage{subfigure,epsfig}
\slugcomment{accepted by ApJ {\it Letters}}

\lefthead{}
\righthead{}

\begin{document}

\title{ Relics of Double Radio Sources }

\author{ K. S. Dwarakanath \& Ruta Kale}
\affil{Raman Research Institute, Bangalore 560 080, India}
\email{dwaraka@rri.res.in, ruta@rri.res.in}

\shorttitle{Radio relics}
\shortauthors{ Dwarakanath \& Ruta}

\begin{abstract}

We have formed a new sample which consists of extended extragalactic radio sources 
without obvious active galactic nuclei (AGN) in them. Most of these sources appear to be 
dead double radio sources. 
These sources with steep spectra ($\alpha < $ -1.8; S $\propto \nu^{\alpha}$)
were identified using the
74 (VLSS) and the 1400 MHz (NVSS) surveys and further imaged using the Very Large Array
(VLA) and the Giant Meterwave RadioTelescope (GMRT). 
The radio morphologies of these sources are rather unusual in the sense that 
no obvious cores
and jets are detected in these sources, but, two extended lobes are detected
in most.  The mean redshift of 4 of the 10 sources reported here is $\sim$ 0.2.
At a redshift of 0.2, the linear extents of the
sources in the current sample are $\sim$ 250 kpc with their spectral 
luminosities at 1.4 GHz in the range 2-25 x 10$^{23}$ W Hz$^{-1}$. 
The steep spectra of these sources is a result of the cessation of 
AGN activities in them about 15 -- 100 million years ago. Before the cessation of AGN
activity, the radio luminosities of these galaxies were $\sim$ 1000 times brighter than
their current luminosities and would have been comparable to those of the brightest
active radio galaxies detected in the local universe 
(L$_{1.4} \sim$ 10$^{27}$ W Hz$^{-1}$).  The dead radio galaxies reported here
represent the $'$tip of the iceberg$'$ and quantifying the abundance of such a population 
has important implications to the life cycle of the AGN.

\end{abstract}


\section{INTRODUCTION}

Sources with steep spectra ($\alpha <$ -1.5; S $\propto$ $\nu^{\alpha}$) at low radio frequencies
($\nu < \sim$ 1 GHz) are interesting to study from many points of view. Such sources could be
(a) pulsars (millisecond, or, otherwise), (b) haloes in galaxy clusters, (c) relics of
double radio sources, or,
(d) high redshift radio galaxies. Although sources of this kind are known, more in some category
than in the other, they are rather rare. Multi frequency studies of these objects are even rarer.
In this background, it is interesting to follow up steep-spectrum sources identified 
from sensitive low-frequency surveys. 

Several studies of steep spectrum sources extracted from various radio surveys  
have been carried out. Based on the radio surveys at 38 and 178 MHz, Baldwin and
Scott discussed sources with spectra steeper than -1.2 and concluded that about half
of these sources are associated with clusters of galaxies (Baldwin \& Scott, 1973).
Komissarov and Gubanov (1994) modeled the evolution of the synchrotron spectrum
of a number of steep-spectrum sources ($\alpha < $ --2.0) found in the centers
of rich clusters. The spectra of radio relics in galaxy clusters were modeled by
Slee et al (2001) and Kaiser and Cotter (2002).
 Gopal-Krishna et al (2005) extracted a sample of
52 sources from 
the Ooty lunar occultation survey, Molonglo 408 MHz surveys MC1, MC2 
 and MIT-Greenbank survey.
The spectral indices of sources in this sample in  the frequency range 300 MHz - 5 GHz
 are steeper than -1.1 but flatter than -1.5. This study detected
many radio sources in the redshift range 0.4 - 2.6, some of them in galaxy clusters.
There appear to be some radio sources in this sample which are at redshifts
even beyond 2.6.  Klamer et al (2006) have studied a sample of 
76 steep spectrum ($\alpha$ in the range -1 -- -1.6)  sources extracted from the 
SUMSS and the  NVSS surveys to find high z radio galaxies 
and advance a plausible cause behind the  z-$\alpha$ correlation. 
Cruz et al. (2006) studied a sample of 68 sources extracted from the 6C(151MHz) survey. 
The spectral
indices (between 151 and 1400 MHz) of these sources are in the 
range -1 -- -1.4 and these sources are confined to 
$\sim 0.4$ sr of the sky. The redshifts of these sources are in the range 0.2 - 3.3.
Parma et al (2007) studied a sample of steep-spectrum ($\alpha < -$ 1.3) 
sources selected by comparing the NVSS and the WENSS catalogs and found six
dying sources and three restarted sources.

Recently, results from the VLA 74 MHz-survey have become available. This survey has a resolution of
$\sim$ 80$''$ with an average rms of $\sim$ 0.1 Jy/beam (Cohen et al 2006).
In the Data Release I the area covered is $\sim$ 6 sr.
This survey is the most sensitive survey to date in this frequency range. This survey
data can be effectively used to identify the steep-spectrum sources. On similar
lines of motivation, a limited area of the sky ($\sim$ 0.05 sr) was analyzed by 
Cohen at al (2004) and further followed by imaging in the near-infrared
by Jarvis et al (2004). The radio sources in their list had spectral
indices (between 74 and 1400 MHz) in the range -1.2 -- -1.8. 
These studies have yielded potentially
interesting candidates for high redshift radio galaxies, relics of double radio
sources, and cluster halo sources.

\section{STEEP-SPECTRUM SOURCES}

In order to select steep-spectrum sources, the Data Release I of 
the VLA 74-MHz survey  (VLSS) and the 1.4 GHz NRAO VLA Sky Survey (NVSS) were used.
The resolutions of the VLSS and of the NVSS surveys are $\sim$ 80$''$ and 45$''$ respectively.
The detection limits of the VLSS and the NVSS surveys are $\sim$ 0.5 Jy/beam and 2 mJy/beam
 respectively corresponding to 5 times the RMS in the respective images.
The source positions in the VLSS survey ($\sim$ 32000 sources) were compared with those
of the NVSS sources ($\sim$ 1.8 million sources).
More than $\sim$ 80 \% of the sources in the VLSS had a counterpart in the NVSS within
$\sim$ 15$''$ -- well within the expected value based on positional uncertainties in the
two surveys.  The spectral indices (flux density $\propto \nu^{\alpha}$)
of all these sources were estimated between 1400 and 74 MHz. The distribution
of spectral indices mimics a Gaussian distribution with a mean spectral index of 
--0.79 and an rms of 0.2.
These values are consistent with those obtained in earlier higher frequency surveys.

For the purposes of this study, $'$steep-spectrum sources$'$ are those which have spectral
indices smaller than, or, equal to -1.8 (mean-5$\sigma$). These are continuum sources with
the steepest spectra that have been studied yet.  There were 38 such sources in the VLSS. The 
NASA Extragalactic Database (NED) and the SIMBAD were searched 
for any other information about these sources.
Of the 38, only 12 of them had non-stellar sources within an arc minute of their
positions.  In addition, there are 3 more sources
which do not have an optical counterpart, but are considered 'extended' based on the 
NVSS data. Given the complementary information
they have, these 15 sources were further followed up with imaging at low-frequencies.

\section{OBSERVATIONS AND DATA ANALYSIS}

Of these 15 sources, two sources 0741+7414 (in ZwCl 0735.7+7421) and 0041-0923 (in Abell 85)
have already been imaged at 330 and 1420 MHz with the VLA (Cohen et al,
2005; Slee et al 2001; Young 2004) with interesting results regarding                  
the nature of these steep spectrum sources.
The remaining 13 sources  were observed at 330 MHz with the VLA in the A configuration ($\sim$ 10$''$ resolution)
during June - Sep, 2007. The observations were carried out in the 4 IF mode with a bandwidth
of 6.25 MHz and 16 channels. The two frequencies were centered at 321.5 and 328.5 MHz respectively.
In each observing session, 3 to 4 sources were cycled through for a
total of 6 to 8 hours giving an integration of time $\sim$ 2 hours on each source, but
with a better visibility coverage. 

Of the 13 sources observed with the VLA, 4 sources had poor quality data from the VLA
to EVLA transition-related problems. They were discarded.
The remaining 9 sources were observed with the GMRT. The observations with the GMRT were carried out
during 5 - 15 May, 2008. Observations were carried out at a center frequency of 1287 MHz with a
16 MHz bandwidth and 128 channels. Two sources were cycled through in an observing session with 
a total duration of $\sim$ 10 hours, giving an integration time on each source $\sim$ 5 hour.

Both the VLA and the GMRT data were analyzed using the Astronomical Image Processing
System. 
The multichannel nature of the continuum data was effective in identifying and
excising the radio frequency interference (RFI). About 10 - 30 \% of the data was lost
to RFI in the VLA and in the GMRT data. 

\section {RESULTS}
 
All the 9 sources except one were detected at both the frequencies. The undetected source was
a result of confusion from a brighter source in the lower-resolution VLSS and NVSS surveys. 
Images of the relevant portions of the 6 (out of 8; to conform to the ${\it Letters}$ length ) 
fields at 328 and 1287 MHz,
 convolved to the common resolution, are displayed in 
Figs. 1-3. The integrated flux densities of the steep-spectrum sources estimated from
these images are given in Table 1. This table has 10 entries since one of the fields
(0128) has three sources with different spectral indices. 
In the case of rest of the sources, the spectral indices of the two lobes
were estimated separately and were found to be in agreement with each other
and with the respective integrated spectral indices within the errors (Table 1). 
Images from the NVSS and VLSS were used to estimate the
corresponding flux densities at 1400 and 74 MHz respectively. The flux density values at 74 MHz 
are given 
in Table 1. Where available, the flux densities of the sources at 326 MHz were also extracted from 
the WSRT catalog. The mean differences between the VLA (at 328 MHz) and the WSRT (at 326 MHz)
flux densities and that between the GMRT (at 1287 MHz) and
the NVSS (at 1400 MHz) flux densities are 14\% and 28\% respectively.
The spectral indices estimated based on the flux densities from the current observations and
from the 74 MHz survey are given in Table 1. The spectra of all the sources in Figs. 1-3 are
displayed in Fig. 4.

\section{DISCUSSION}

A striking feature of the morphologies of the steep-spectrum sources is their
double-lobed nature in most cases.  The morphologies
of the steep-spectrum sources observed here are similar to those of radio galaxies 
rather than that of halos and relics observed
in galaxy clusters. Furthermore, there are no clear indications of cores and jets
in any of these sources (with the exception of 0128B) even in the highest resolution
($\sim$ 4$''$) GMRT images at 1287 MHz.
 The morphologies, the absence of cores and jets
 and the steep spectra of these sources
imply that they are most likely fossil radio sources. There is no evidence
that any of these sources reside in galaxy clusters except perhaps the sources
1152 and 2216.  In contrast to this, both
the steep-spectrum sources from the VLA 74 MHz survey imaged in recent 
times turned out to be in galaxy clusters (Slee et al 2001; Cohen et al 2005).

Four of the ten sources are at a redshift of $\sim$ 0.2 (see notes to Table 1). 
For the purposes of this discussion we will assume the current sample
of steep-spectrum sources to be at a redshift of 0.2. This is not a serious
drawback since the conclusions we will draw on the nature of these sources
is not sensitive to this assumption.  The implied range of radio 
luminosities (H$_{o}$ = 75 km s$^{-1}$ Mpc$^{-1}$) at 1400 MHz is 
2-25 $\times$ 10$^{23}$ W Hz$^{-1}$, close to the
value where the distribution of local AGNs peak (Condon 1989, Sadler et al 2002). 
The linear extents of the steep-spectrum sources are in the range 100 - 400 kpc.
The equipartition magnetic fields in these systems
are in the range 2 - 10 $\mu$G. 

Several previous studies of spectral evolution and modeling exist to
account for steep spectra
(Komissarov \& Gubanov 1994, Goldshmidt \& Rephaeli 1994, Slee et al 2001,
Ensslin \& Gopal-Krishna 2001). The basic synchrotron theory is
described in Kardashev (1962) and Pacholczyk (1970). Spectral steepening
in extragalactic radio sources is due to differential 
energy loss of the relativistic electrons as a function of their energy (E).
The energy losses relevant for the sources of interest are synchrotron, 
inverse-Compton and adiabatic losses. Synchrotron losses are proportional
to B$^{2}$E$^{2}$ while the inverse Compton losses are proportional
to B$_{ic}^{2}$E$^{2}$, where, B is the magnetic field strength in the
source and B$_{ic}$ (= 3.25$(1+z)^{2} \mu$G) is the magnetic field strength with energy density 
equal to that of the cosmic microwave background. The adiabatic losses
are connected to the change in the volume of the source. 
Ensslin and Gopal-Krishna (2001) set up a formalism in which they start
with the equation relating the change in the energy of the relativistic
electrons with the losses mentioned above. They further assume 
sufficient pitch angle scattering to keep the electron pitch angle
distribution isotropic (JP model; Jaffe \& Perola 1973).  The solution to this equation
leads to the expression for the synchrotron luminosity at any given
frequency, L$_{\nu i} = c_{3}B_{i}V_{i}\int\limits_{pmini}^{pmaxi}\!dpf_{i}(p)F(\nu/\nu_{i}(p))$ 
(eq. 19 in their paper). In this equation, $c_{3}$ is a constant and 
$F(\nu/\nu_{i}(p))$ is the dimensionless spectral emissivity of a 
mono-energetic isotropic electron distribution in isotropically oriented
magnetic fields. The symbols $B_{i}$ and
$V_{i}$ are the magnetic field and the volume of the source respectively
 in the i'th phase.
In the current study, the two phases considered are the injection and 
the expansion phases. Starting from an initial power law, the resulting electron
spectrum (due to the losses mentioned earlier) in the i'th phase is given by $f_{i}(p)$. 
This equation does not have an analytical
solution and was evaluated numerically in the current study.
A more detailed discussion of this formalism 
is beyond the scope  of this {\it Letter} (see Ensslin \& Gopal-Krishna (2001)
for details).

The above formalism was used to estimate the model spectra that best-fit the
observed data points. 
The input parameters to the computation of the model spectra are 
(a) the current volume of the source, (b) the source magnetic field 
and (c) the initial index $\gamma$ in the
energy distribution of the relativistic electrons (N=N$_{o}$E$^{-\gamma}$).
The spectra were obtained at any given time, t = t$_{CI}$ + t$_{RE}$, where t$_{CI}$ 
is the duration of the continuous injection phase and 
t$_{RE}$ is the duration of the relic phase during which the injection
of relativistic electrons is switched off (see Slee et al 2001).
It is possible to estimate the parameters (viz. t$_{CI}$, t$_{RE}$, and
B) for each source that best-fit the
respective spectra. 

The current volume of the source
was estimated for a linear size of 250 kpc. This linear size corresponds
to an angular size of $\sim$ 1$'$ (see Fig. 1-3) at a redshift of 0.2 
(H$_{o}$ = 75 km s$^{-1}$ Mpc$^{-1}$). 
A value of $\gamma$ = 2.5 was assumed corresponding to an $\alpha$ = --0.75,
where the flux density is $\propto \nu^{\alpha}$. For different sources,
a range of t$_{CI}$'s,
 t$_{RE}$'s and B's  were considered and the model spectra obtained. 
The model spectra were then compared with the observed data. The model
spectrum with a single value of B has an exponential high frequency cut-off
and is unable to fit the observed data. This is a well-known problem
as was noted by Komissarov \& Gubanov (1994) and Slee et al (2001).
However, if the spectra for a range of field strengths is obtained and
combined, the sharp JP cutoff is smoothed out and the integrated spectrum
 fits the data well. Such a model is considered more physical due to the
different strengths of magnetic fields believed to exist in any of these systems
(Slee et al 2001).  The solid
lines in Fig. 4 indicate the best-fit model spectra to the corresponding 
data points. The model spectra in each case were anchored to the flux
density at 74 MHz to fix the y-normalization. 
The dotted and dashed lines in Fig. 4 indicate the 
model spectra when t$_{RE}$ was changed by $\pm$ 30\% respectively with respect
to the best-fit values quoted in each panel. It is evident from these plots
that a $\pm$ 30\% change from the best-fit value of t$_{RE}$ produces model
spectra which are in disagreement with the observed data points and 
predicts values beyond the 3$\sigma$ errors of the measurements.
So, the errors in the estimates of t$_{RE}$'s are not more than $\pm$ 30 \%.
Similar considerations indicate that the errors in the estimates of t$_{CI}$'s
are not more than $\pm$ 50 \%. Although there are only three data points, 
the shape of the spectrum is well-constrained since the data points are
well-separated in frequency. Since the shape of the spectrum is a 
sensitive function of t$_{CI}$, t$_{RE}$ and B it has been
possible to estimate these parameters. The ranges of parameters that account for the 
observed spectra of the current sample of sources
are the following : 10 $<$ t$_{CI}$ $<$ 400 My,
15 $<$ t$_{RE}$ $<$ 90 My and 2 $<$ B $<$ 10 $\mu$G. 
The mean value of the ratio t$_{RE}$/t$_{CI}$ for the current sample
of sources (Table 1) is 1.3.
This mean value of t$_{RE}$/t$_{CI}$ implies that most of the sources
discussed here are at a stage when the central engine has been off for as long
as it was on. During the
computation of model spectra, it is also possible to keep track of
the relative decrease in the radio luminosity as a function of time. This
ratio varies in the range 10$^{2}$ to 10$^{4}$ depending on 
the value of t$_{RE}$/t$_{CI}$ for different sources.
Since the current radio luminosities of these sources are $\sim$ 10$^{24}$
W Hz$^{-1}$ at 1.4 GHz, the peak radio luminosities of these sources
when the AGN was active would have been $\sim$ 10$^{27}$ W Hz$^{-1}$, brighter
than the brightest of the currently known radio galaxies. 
While the current sample represents sources when the relic phase is 
comparable to the active phase, the Parma et al (2007) sample represents radio galaxies 
when the relic phase is an order of magnitude smaller than the active phase.

The detection limits of the surveys at 1400 and 74 MHz are 2 and 500 mJy/beam
respectively. The sensitivities of these two surveys are comparable for sources
with a spectral index of -1.8. Hence, such steep-spectrum sources detected at 1400 MHz
would also be detected at 74 MHz.  The peak luminosities of the sources 
discussed here is L$_{1.4} \sim$ 10$^{27}$ W Hz$^{-1}$, when they were active.
Most of these sources are close to the detection limits of the two surveys
used to identify them. However, the luminosity
function of currently active AGNs indicates that their number density 
is about 100 times
higher at L$_{1.4} \sim$ 10$^{24}$ W Hz$^{-1}$ (Sadler et al 2002) than at 
L$_{1.4} \sim$ 10$^{27}$ W Hz$^{-1}$. At the corresponding ages of the sources
reported here, majority of the currently active sources (with L$_{1.4} \sim$ 10$^{24}$ W Hz$^{-1}$)
will remain undetected at the current sensitivity limits of low-frequency surveys.
Hence, the dead radio sources
discussed here represent just the $'$tip of the iceberg$'$ with a large population
of such sources to be unearthed from more sensitive low frequency surveys.
Such surveys ought to detect a large number of AGNs at different stages of their activities (and
hence different spectral indices) and lead to an understanding of their
workings and evolution.

\acknowledgements 

We thank Biman Nath and Ravi Subrahmanyan for critical comments
and discussions.
The National Radio Astronomy Observatory is a facility of the 
National Science Foundation operated under cooperative agreement 
by Associated Universities, Inc. 
GMRT is run by the National Centre for Radio Astrophysics of the 
Tata Institute of Fundamental Research.

\clearpage

\begin{figure}[!t]
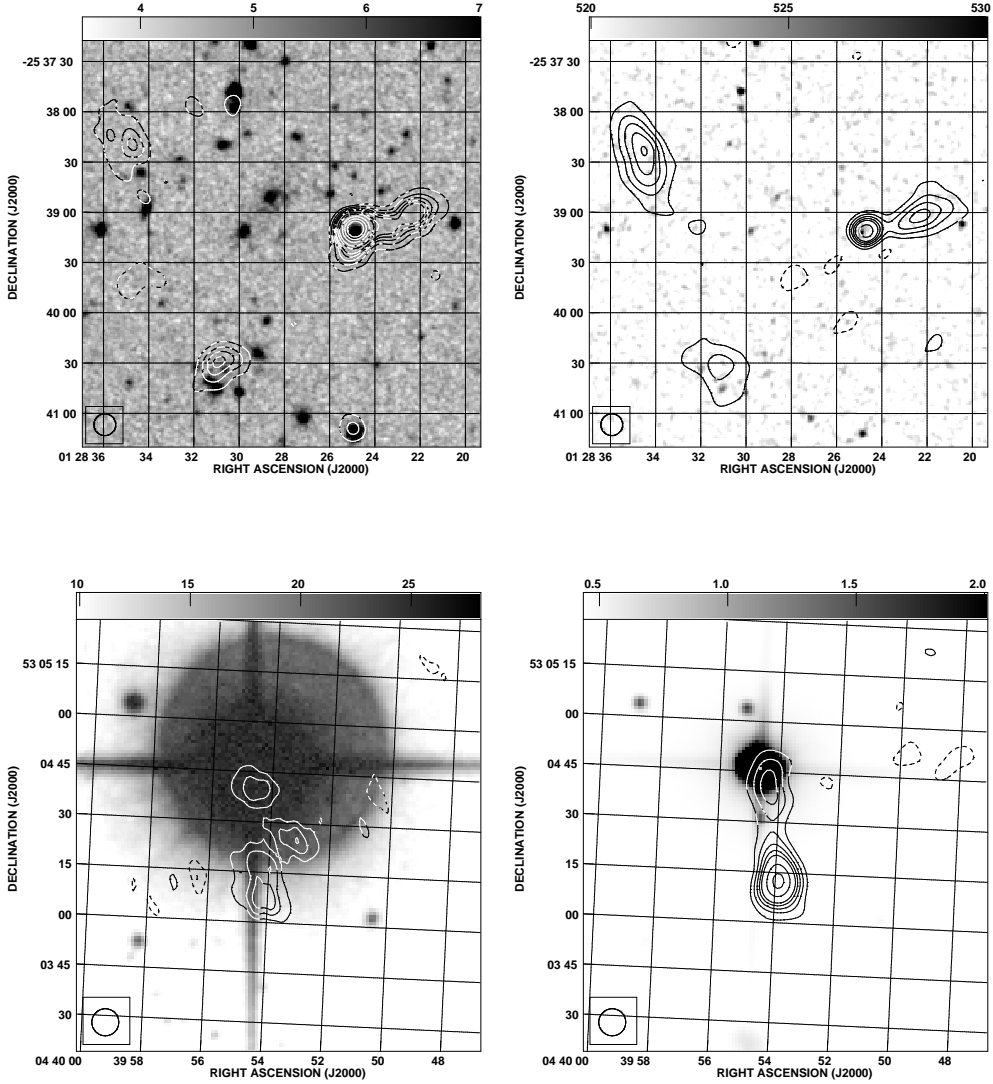

\centering
\epsfig{figure=0128.PS2,width=0.4\linewidth}
\epsfig{figure=0128.PS1,width=0.4\linewidth}\\
\epsfig{figure=0439.PS2,width=0.4\linewidth}
\epsfig{figure=0439.PS1,width=0.4\linewidth}
\caption{{\it top (0128)} : GMRT (left, 1287 MHz) and VLA (right, 328 MHz) radio images (contours) 
overlaid on optical and 2$\mu$ All Sky Survey images (in grey) respectively.
The synthesized beam is 13$^{''}$ X 13$^{''}$ and the rms values are 0.14 (left) and 3.5 (right)
mJy/beam. 
The contours are at -0.5, 0.5, 1, 1.5, 2, 2.5 and 3.75 (left) and at -8, 8, 16, 24, 32, 40,
60, 80, 120, 160 and 200 (right) mJy/beam.
The source to the NE is 0128A, to the NW is 0128B and to the south is 0128C.
{\it bottom (0439)} : The synthesized beam is 8$^{''}$ X 8$^{''}$.
 The rms values are 0.14 (left) and 1.6 (right) mJy/beam.
The contours are at -0.5, 0.5, 1 and 1.5 (left) and 
at -4, 4, 8, 12, 16, 20, 30 and 40 (right) mJy/beam.
\label{fig1}}
\vspace{-5pt}
\end{figure}

\begin{figure}[!t]
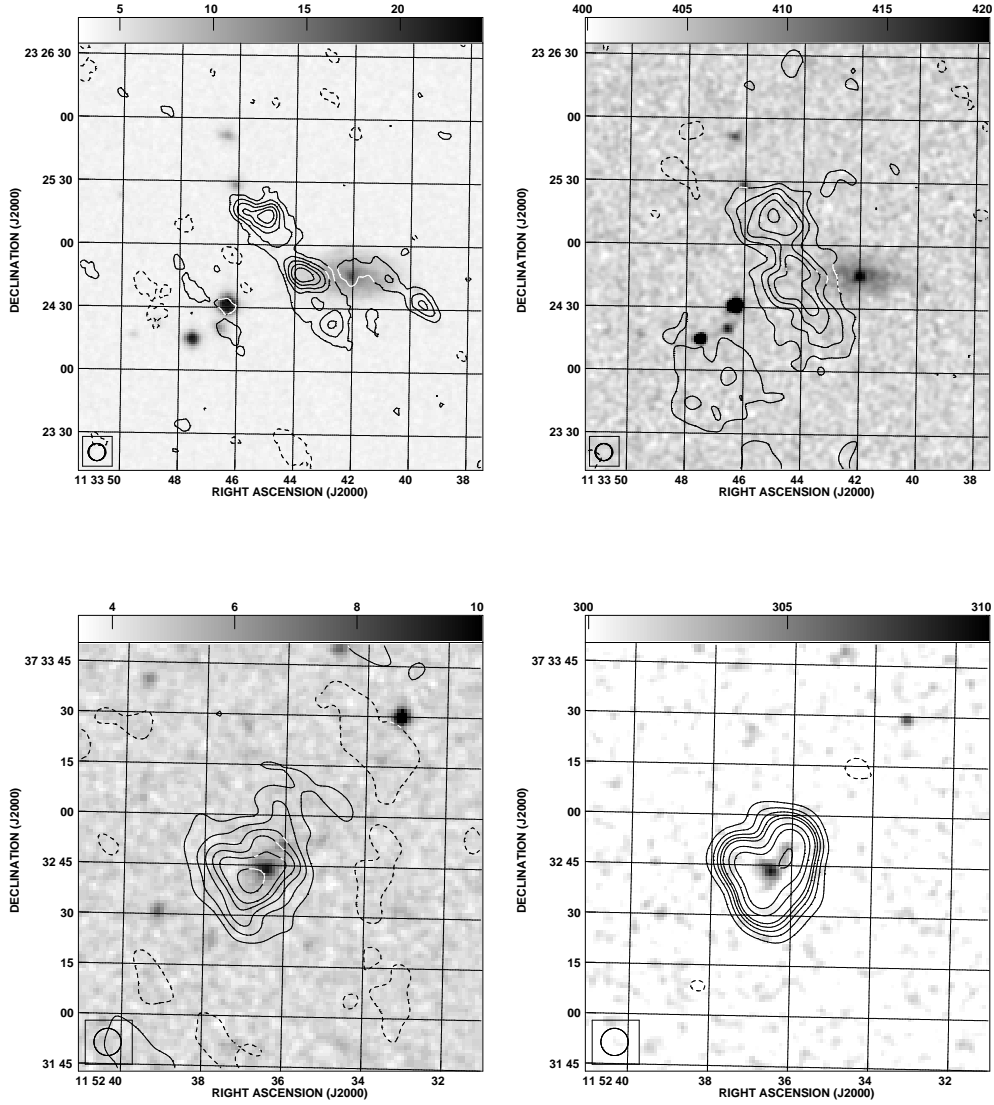

\centering
\epsfig{figure=1133.PS2,width=0.4\linewidth}
\epsfig{figure=1133.PS1,width=0.4\linewidth}\\
\epsfig{figure=1152.PS2,width=0.4\linewidth}
\epsfig{figure=1152.PS1,width=0.4\linewidth}
\caption{
{\it top (1133)}:
The synthesized beam is 8$^{''}$ X 8$^{''}$.
The rms values are 0.1 (left) and 0.95 (right) mJy/beam.
The contours are at -0.2, 0.2, 0.4, 0.6, 0.8 and 1 (left) and 
at -2, 2, 4, 6, 8 and 10 (right) mJy/beam.
{\it bottom (1152)} : The synthesized beam is 8$^{''}$ X 8$^{''}$.
 The rms values are 0.3 (left) and 1.3 (right)
mJy/beam. The contours are at -0.6, 0.6, 1.2, 1.8, 2.4, 3 and 4.5 (left) and at -4, 4, 8, 12, 16,
20, 30, 40 and 60 (right) mJy/beam. 
\label{fig2}}
\end{figure}

\begin{figure}[!t]
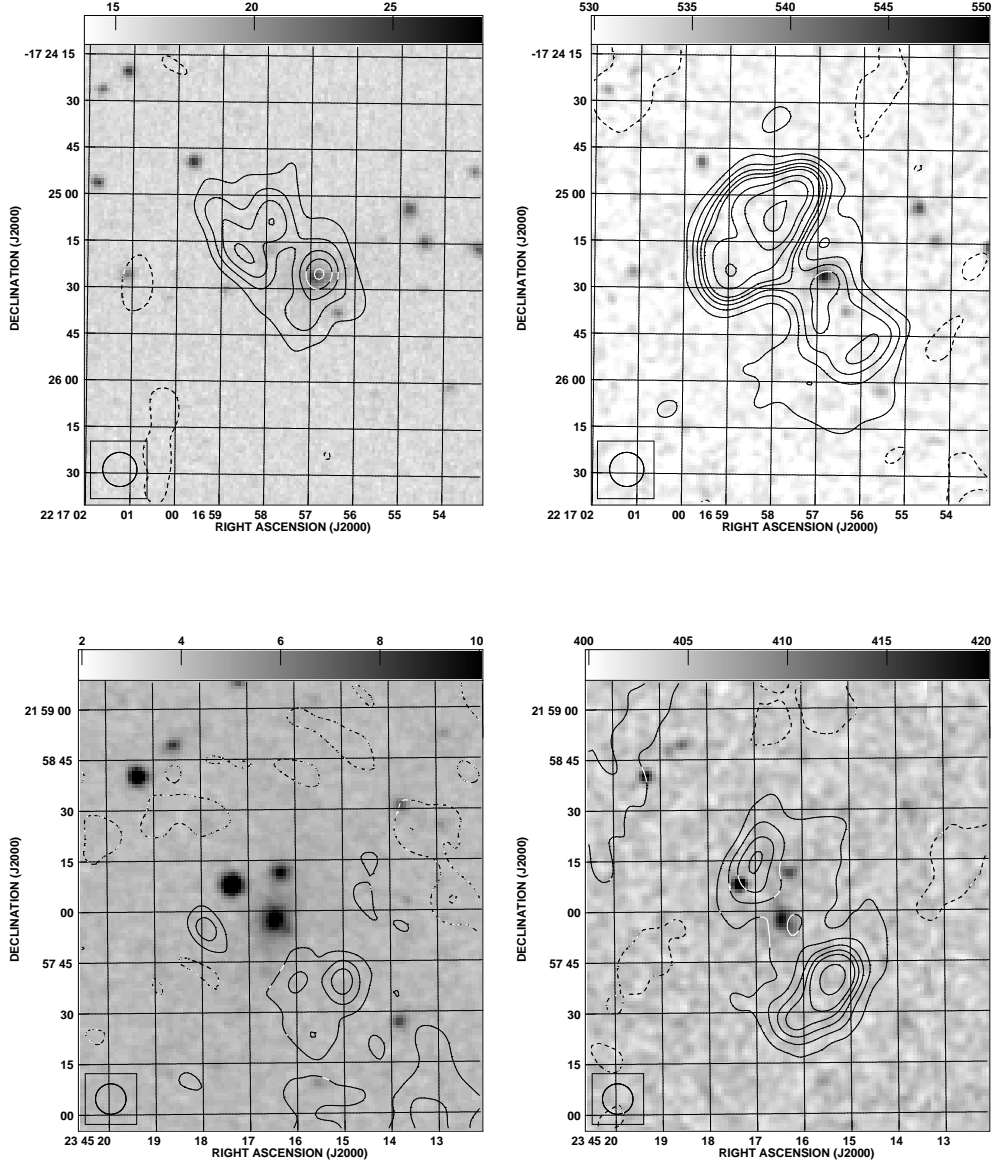

\centering
\epsfig{figure=2216.PS2,width=0.4\linewidth}
\epsfig{figure=2216.PS1,width=0.4\linewidth}\\
\epsfig{figure=2345.PS2,width=0.4\linewidth}
\epsfig{figure=2345.PS1,width=0.4\linewidth}
\caption{ 
{\it top (2216)} :
The synthesized beam is 11$^{''}$ X 11$^{''}$.
The rms values are 0.18 (left) and 2.7 (right)
mJy/beam. The contours are at -0.4, 0.4, 0.8, 1.2, 1.6 and 2.0 (left) and 
at -4, 4, 8, 12, 16, 20,
30, 40 and 60 (right) mJy/beam.
 {\it bottom (2345)} :
The synthesized beam is 9$^{''}$ X 9$^{''}$.
The rms values are 0.44 (left) and 3.0 (right)
mJy/beam. The contours are at -0.7, 0.7, 1.4 and 2.1 (left) and 
at -5.6, 5.6, 11.2, 16.8, 22.4, 28 and 42 (right) mJy/beam.
\label{fig3}}
\end{figure}

\begin{figure}
\figurenum{4}
\epsscale{0.7}
\plotone{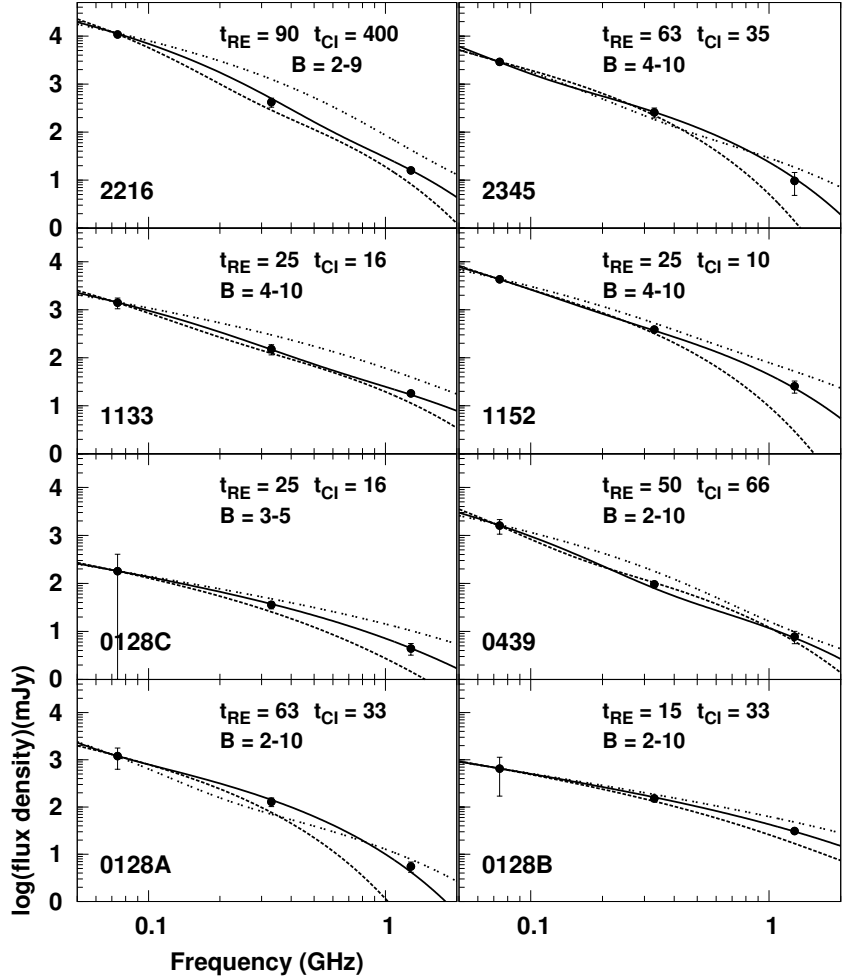}
\caption{ Spectra of all the sources in Figs. 1-3. The solid points are observed
data along with 3$\sigma$ error bars. The solid line is the best-fit model
spectrum. The best-fit parameters are indicated in each panel. The values
of t$_{RE}$ and t$_{CI}$ are in million years and of B are in $\mu$G.  
The dotted
and dashed lines are model spectra when t$_{RE}$ is changed by $\pm$ 30\% 
respectively with respect to the best-fit value. The 3$\sigma$ errors are on 
the t$_{CI}$ values are $\pm$ 50\%.
Note that the model spectra for each source are anchored to the respective data 
value at 74 MHz to fix the y-normalization.
\label{fig4}}
\end{figure}

\clearpage
 


\begin{deluxetable}{ccccccccc}
\tabletypesize{\footnotesize}
\tablenum{1}
\tablecolumns{10}
\tabcolsep 0.05 in
\tablewidth{0pc}
\tablecaption{GMRT \& VLA Observations}
\tablehead{
\colhead{Source}&\colhead{R.A. (J2000)}&\colhead{Dec. (J2000)} 
&\colhead{S74} &  \colhead {S328} & \colhead {S1287}
& \colhead {$\alpha_{328}^{74}$} & \colhead { $\alpha_{1287}^{328}$}
& \colhead{t$_{RE}$/t$_{CI}$}
\\
\colhead{}&\colhead{hh mm ss}&\colhead{o ~~$'$ ~~$''$} 
&\colhead{(VLSS)} & \colhead{(VLA)} & \colhead{(GMRT)}
& & &
}
\startdata
0128A & 01 28 34.551 & -25 38 22.90 & 1200(190) &  129(9) & 5.5(0.45) &  -1.5(.15) & -2.3(.15) & 1.9 \nl
0128B & 01 28 24.697 & -25 39 11.15 & 650(160) & 152(7.5) & 31.0(0.3) &  -1.0(.2) & -1.2(.05) & 0.5 \nl
0128C & 01 28 31.112 & -25 40 31.80 & 180(75) &  35.8(2.1) & 4.4(0.4) &  -1.1(.34) & -1.5(.15) & 1.6 \nl
0439 & 04 39 53.930  & 53 04 12.02 & 1600(180) & 95.6(4.6) & 7.8(0.75) & -1.9(.1) & -1.8(.14) & 0.8 \nl
1114 &  11 14 13.194 & 15 19 44.12 & 600(101) &  43.3(6) & 2.1(0.46) & -1.8(.2) & -2.2(.27) & 0.4 \nl
1133 &  11 33 45.045 & 23 25 14.05 & 1400(118) & 150.8(12) & 18.1(0.40) & -1.5(.1) & -1.6(.09) & 1.6 \nl
1152 & 11 52 36.104  & 37 32 46.63 & 4300(102) & 387(11) & 25.5(2.4) &  -1.6(.03)& -1.9(.08) & 2.5 \nl
2216 & 22 16 58.007  & -17 25 08.03 & 10800(380) &  420(30) & 15.9(0.7) &  -2.2(.07) & -2.4(.09) & 0.2 \nl
2313 & 23 13 46.837  & 38 42 16.13 & 1800(100) & 245(12) & 28.7(8.5) & -1.3(.08) & -1.6(.25)& 1.5 \nl
2345 &  23 45 15.392 & 21 57 39.82 & 2900(114) & 257(20) & 9.6(1.6) & -1.6(.08) & -2.4(.18)& 1.8\nl
\enddata
\tablecomments { All flux densities are integrated values and in mJy. Values within
parentheses are 1$\sigma$ error estimates. The sources with
optical identifications 
are 0128B (2dFGRS S149Z127, z=0.207), 1152 (SDSS J115236.45+373243.7, z=0.229), 
2216 (RBS 1842, z=0.136), and 2345 (2MASX J23451645+2157578, z$\sim$ 0.15 (based on the k-z relation
of van Breugel et al (1999))).
 } 
\end{deluxetable}

\end{document}